\documentclass[twocolumn]{emulateapj}

\usepackage{epsfig}
\usepackage{amssymb}

\bibpunct{(}{)}{;}{a}{}{,}

\newcommand{\msun}{\mbox{$M_\odot$}}
\newcommand{\rsun}{\mbox{$R_\odot$}}
\def\be{\begin{eqnarray}}
\def\ee{\end{eqnarray}}
\def\lsim{\mathrel{\rlap{\lower3pt\hbox{\hskip1pt$\sim$}}
     \raise1pt\hbox{$<$}}} 
\def\gsim{\mathrel{\rlap{\lower3pt\hbox{\hskip1pt$\sim$}}
     \raise1pt\hbox{$>$}}} 

\shorttitle{SN2008D: Repetition of SN1987A}
\shortauthors{}

\begin{document}

\title{Supernova 2008D:  A Repetition of Supernova 1987A In a Binary}

\author{Gerald E. Brown}
\affil{Department of Physics and Astronomy,
               State University of New York, Stony Brook, NY 11794, USA.}
\email{GEB: gbrown@insti.physics.sunysb.edu}

\and

\author{Chang-Hwan Lee}
\affil{Department of Physics, Pusan National University,
              Busan 609-735, Korea.}
\email{CHL: clee@pusan.ac.kr}


\begin{abstract}
The Supernova 2008D is similar to that of SN 1987A without the H-envelope.  \citet{Sod08} reported the serendipitous discovery of the SN2008D at the time of the explosion, accompanied by an X-ray outburst XRF080109.  The central remnant, which we believe to be the black-hole (BH) central engine in the Blandford-Znajek mechanism, is estimated, on the basis of the $7\%$ $^{56}$Ni production, to have a mass of $1.6-1.8\msun$.  This is not much larger than the upper limit of $1.56\msun$ for the mass of the compact object in SN1987A found by \citet{Bet95}; also, on the basis of the $7.5\%$ $^{56}$Ni production, they interpreted it as a low-mass BH.
Redoing the light curve so as to take into account the absence of convective carbon burning from zero age main sequence (ZAMS) $18-24\msun$ and replacing it by carbon shell burning, we see that the remnant in SN2008D must be less massive than in SN 1987A; there of $\sim 18\msun$.  Thus, the maximum neutron star mass is $\sim1.5\msun$.
Note that the metallicity of the host galaxy of SN2008D is similar to that of our Galaxy.
\end{abstract}



\section{Introduction}\label{intro}

\citet{Bro07} showed that the subluminous gamma-ray bursts (GRBs) came from the soft X-ray transient sources; i.e., from BH binaries.  They also showed \citep{Bro08} that LMC X$-$3 could be a relic of a cosmological GRB.  Therefore, we consider a binary as  a possible progenitor for XRF080109/SN2008D.  We also compare the explosion in SN1987A with the explosion in SN2008D.  The latter is, of course, that of a helium star, because the hydrogen envelope was removed in common envelope evolution.
Our Case C mass transfer, which requires the He in the He star to be burned {\it before} common envelope evolution is helpful in preserving the binary preceding collapse.  In fact the He-core burning must have taken place long preceding collapse.

ZAMS stars of the mass we consider here go first into a neutron star which then goes into a BH in a delayed explosion \citep{Tim96}.

Much of the H envelope in the progenitor of SN1987A was present at the time of the explosion, but this is shown to not affect the later evolution into a BH.

In section~\ref{BZM} we discuss the application of the Blandford-Znajek mechanism as central engine for GRBs.  In section~\ref{BH87A} we  discuss hypercritical accretion which was necessary in the evolution of the compact object in SN1987A.  In section~\ref{Expl} we construct the explosion and in section~\ref{Core} we compare the core masses of SN1987A and SN2008D. In section~\ref{GRBs} we discuss the relation to other GRBs.

\section{Blandford-Znajek Mechanism}\label{BZM}

\citet{Bro07} evolved the subluminous GRBs as the soft X-ray transient sources; namely they came from binaries.  In the case of XRF060218 , the donor (secondary star) was estimated to have a mass of $m_D \approx 10\msun$, scaling from Cyg X$-$1 \citep{Bro08}.
On the other hand, we shall show that the mass of the SN2008D progenitor is probably about the same as that for SN2006aj. The former is, however, in a galaxy of roughly solar metallicity whereas the galaxy of the latter has low metallicity.

What interests us are: a) Both of the binaries have little $^{12}$C, which tells us that both progenitors came from near the threshold for BH formation, ZAMS $18\msun$ (see fig.1 of \citet{Bro01}). b) Both XRFs are subluminous.  The ram pressure energy in clearing the ambient matter out of the way for $\gamma$-rays, cuts down their energy substantially.

We discussed the Blandford-Znajek mechanism in which the rotating BH was the central engine in powering the subluminous GRBs in \citet{Bro07}.
We use the XRF lifetime (eq.~(8) in \citet{Lee00})
\be
\tau_{\rm BZ}=2.7\times10^3(10^{15}{\rm G}/B)^2(\msun/M_{\rm BH})\ {\rm s}.
\ee
By equating $\tau_{\rm BZ}$ with the $T_{90}=2100\pm100$ s of XRF060218 we find $B_{15}\cong1.3$ G, and in the case of XRF080109 with $T_{90}=600$ s \citep{Tan08} we find $B_{15}=1.6$ G, both reasonable.

\begin{figure}
\centerline{\epsfig{file=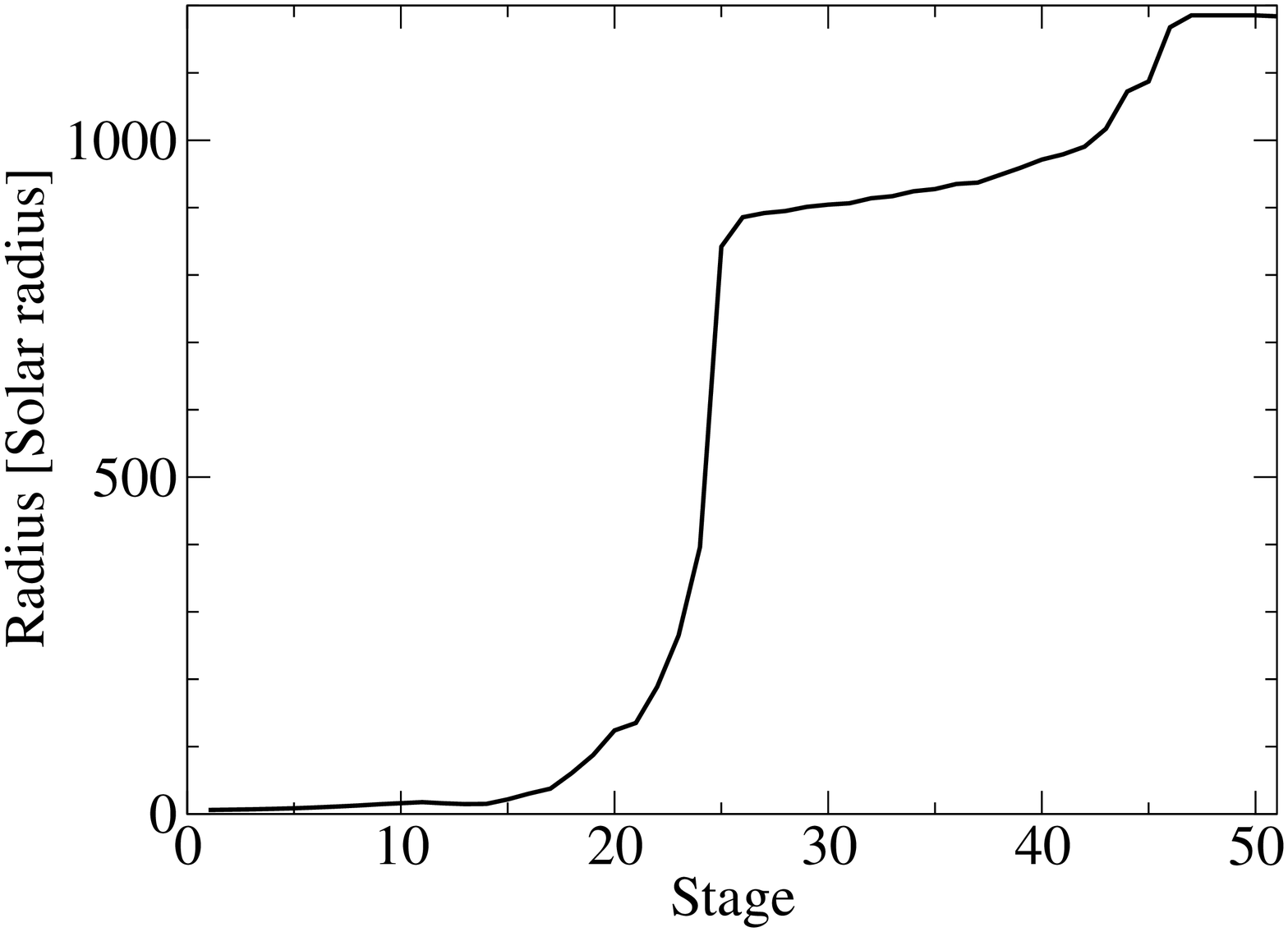,height=2.5in}}
\caption{The radial evolution of \citet{Sch92} 20 $\msun$.  The rapid rise from 1,000 to $1,200\rsun$ is the time of the He shell burning; i.e., of He red giant.}
\label{fig1}
\end{figure}

In Fig.~\ref{fig1} we show the ZAMS $20\msun$ from \citet{Sch92}.  If we use Case C mass transfer, the core He must be burned and then the He envelope goes into He-shell burning and the supergiant expands.  The donor, the secondary star in the binary, is of relatively low mass compared with the supergiant.  The Roche lobe overflow can start only after He core burning is finished; otherwise Case A or B mass transfer takes place.  The donor mass is small compared with that for the giant, so its Roche lobe will only be $\sim20\%$ of that of the giant, with Roche contact established when the giant begins He-shell burning; i.e. the initial separation will be $a_i\sim1200\rsun$ \citep{Bro01b}.

Now, initially, we expect the hypernova heat and XRF (kinetic) energies to be about equal, from equipartition.  However, the X-rays have to produce ram pressure work to clear the way through the He star for the jet, and this takes up most of the kinetic energy.

The heat energy, which is transformed into kinetic energy, of the supernova is about 5 bethes for SN2008D and 2 bethes for SN2006aj.  Using \citet{Lee02} we can find the mass of the donor (secondary star) that will give SN2008D $2.5$ times more angular momentum energy than SN2006aj.

Using the relation between the He core to the star
$ M_{\rm He}=0.08(M_{\rm Giant}/\msun)^{1.45}\msun $
\citet{Lee02} found that the separation following common envelope evolution was
\be
a_f=\left(\frac{M_d}{\msun}\right)\left(\frac{M_{\rm Giant}}{\msun}\right)^{-0.55}a_i
\ee
where $M_d$ is the donor (secondary star) mass.  From Kepler we have the preexplosion period
\be
\frac{days}{P_b}=\left(\frac{4.2\rsun}{a_f}\right)^{3/2}\left(\frac{M_d+M_{\rm Giant}}{\msun}\right)^{1/2}
\ee
Since we take $M_{\rm Giant}$ and $M_{\rm He}=4.6\msun$ to be the same for SN2006aj and SN2008D (see Sec.~\ref{Expl}), we can easily estimate the ratio of orbital rotation rates ($\Omega_{\rm orbital}$):
\be
\frac{\Omega_{2008D}}{\Omega_{2006aj}}=
\left(\frac{M_{d,2008D}+M_{\rm He}}{M_{d,2006aj}+M_{\rm He}}\right)^{1/2}
\left(\frac{M_{d,2006aj}}{M_{d,2008D}}\right)^{3/2}.
\ee

The extractable spin energy in Blandford-Znajek is
\be
E_{\rm rot} \propto f(a_\star)M_{\rm BH}c^2
\ee
where $a_\star$ is the dimensionless spin parameter given by
$a_\star = J_{\rm BH} c/GM_{\rm BH}^2 \approx I_{\rm He}\Omega_{\rm He} c/GM_{\rm BH}^2$
and
\be
f(a_\star) = 1 - \sqrt{\frac{1}{2}(1+\sqrt{1-a_\star^2})}.
\ee
Here we assume that the angular momentum is conserved during the collapse from He star to BH, neglecting the (small) mass left in the disk, some of which will be lost in the explosion. 
For low $a_\star$, $f(a_\star) \cong a_\star^2 /8$. We therefore take the energy as proportional to $\Omega_{\rm He}^2$.
By assuming the tidal locking between the orbital period and the spin period of He star \citep{Lee02}, one can take $\Omega_{\rm He} \approx \Omega_{\rm orbital}$.
  This, then, gives a donor mass of $\sim 6.8\msun$ for SN2008D.

\section{SN1987A Evolved Into A Black Hole}\label{BH87A}

It is now more than 20 years since the explosion of SN1987A.  Nothing has been seen of it.  \citet{Bro94} wrote down an analytical scenario showing that the compact object went into a BH, the mass of which \citep{Bet95} was shown to be $\leq1.56\msun$ from the $7.5 \%$ $^{56}$Ni production.  This calculation involved a $3900\pm400$ km separation distance (between matter that fell back and matter that went to infinity), but the density was low at that point, so the BH mass was insensitive to the separation.  The assumption of \citet{Bet95} that the compact object was a BH was strengthened by the \citet{Che89} work,
which showed that after only $\sim1$ year SN1987A should be observed with a luminosity $\sim L_{Edd}=4\times10^{38}$ ergs s$^{-1}$ were a neutron star present, whereas the actual luminosity was $4\times10^{36}$ ergs s$^{-1}$, two orders of magnitude less.  A BH was calculated to contribute at the level of $10^{34}-10^{35}$ ergs s$^{-1}$ \citep{Bro94}.  The difference could come from radioactivity of the clouds.

Hypercritical accretion; i.e., accretion at a rate faster than Eddington, was employed in the \citet{Bro94} work as well as by \citet{Che89}.  The excess radiation energy was carried off by neutrinos.

Recently a clear proof that hypercritical accretion was necessary in the evolution of the binary M33 X$-$7 was given by \citet{Mor08}.  The present-day binary with period $3.45$ days would have broken up by loss of mass soon after the explosion which formed the BH had hypercritical accretion not taken place.  Hypercritical accretion into the BH of M33 X$-$7 was required.

In many cases, hypercritical accretion may not be possible because of the large amount of angular momentum which must be present, or other reasons, but it is clear that in some cases hypercritical accretion is possible.  In fact, many cases of Bondi accretion do involve hypercritical values, as in the \citet{Bro94} work.

\section{The Explosion}\label{Expl}

Now, the common envelope evolution in the binary progenitor of SN2008D must, like those of other GRBs \citep{Bro07}, be of type I$_c$; i.e., the He burning must be finished before the hydrogen envelope is removed.  We have shown that the BH results from an $\sim18\msun$ ZAMS giant.  However, we clearly do not see the explosion until after the core He is burned.  From \citet{Bro01}, we see that convective carbon burning ceases just at the threshold $18\msun$ ZAMS mass for BH evolution, basically just at the giant progenitor mass of SN1987A.  There is carbon present from shell burning, at about $10\%$ relative abundance, which is substantially hotter than it would be were it burned convectively.  The precise abundance is, however, uncertain because the shell burning is chaotic.

The He star corresponding to a ZAMS $18\msun$ star has a mass of $\sim4.6\msun$, somewhat less than the \citet{Tan08} value of $6\msun$.  But whereas \citet{Tan08} finish the maximum in the light curve off by a ``little bit of carbon", we believe that, since there is not enough carbon to burn convectively, a small amount of the hotter shell that burned carbon will do.  The helium will first be seen with the advent of He shell burning; i.e., as the helium burning goes red giant.  Just this low-mass He star will expand to several $\rsun$ ending in the ignition of more massive metals \citep{vdH92}.  So far the evolution should be similar to that of SN2006aj, which also showed a lack of carbon lines and also came from an $\sim18\msun$ giant \citep{Bro07}.  In the case of SN2006aj a shell of oxygen comprising $\sim0.1\msun$ and expanding at velocities between 20,000 and 30,000 km/s may be responsible for the early ultraviolet brightening \citep{Maz08}.  Oxygen is produced in the He rich layer, but \citet{Tan08} find a fraction of only $0.01\msun$.

As noted earlier, with a secondary star of $\sim 6.8 \msun$, SN2008D will be substantially more energetic than SN2006aj, with an estimated secondary star mass of $\sim10\msun$ \citep{Bro08}, which is the same as the Galactic XTE J1819$-$254 or GRS1915$+$105 at the time of the explosion \citep{Bro07}.  These would have given subluminous GRBs, just as found for SN2006aj.  However, the $6.8\msun$ donor of XRF080109 is closer in mass to the $\sim4\msun$ in LMC X$-$3 \citep{Bro08}, which was found to be a relic of a cosmological type GRB.  As noted in \citet{Bro07}, donors of about this mass should give the maximum rotation energy that can be accepted.  The kinetic energy which feeds the GRB in XRF080109 should be of the same general size as the energy of that of SN2006aj, from equipartition.  However, the XRF must pay ram pressure energy to clear the passage of material through the He star so that $\gamma$-rays can be emitted.
But we estimate that there would still be enough energy so that the explosion would be luminous, as seen.

In summarizing the initial X-rays, these come similarly to those in XRF060218, but for only $1/4$ to $1/3$ of the time.

The hypernova takes place much later, at $\sim19$ days following the XRF.  The general scenario is that the hypernova is powered by the deposition of BH rotational energy onto the accretion disk.  The viscous timescale (\cite{Bet03}, p.359)
\be
\tau_{visc}=\frac{1}{\alpha}\left(\frac{r}{h}\right)^2 \tau_{dyn},
\ee
where with the strong ($\sim10^{15}$ G) magnetic field perpendicular to the accretion disk the $\alpha$ in the $\alpha$-viscosity will not be much less than unity, but the inner disk is quite thin, $h\sim0.2r$ \citep{Mac99}.  The dynamical time for direct collapse would be only $\sim12$ s, giving $\tau_{visc}\sim1$ hour.  However, the delivery of heat from the BH to the accretion disk is like the little ball bouncing around in a roulette wheel, the sides of which are the accretion disk.  Only ocasionally will the ball hit the center of the disk and it is easy to see that the $\tau_{visc}$ could be lengthened by two or three orders of magnitude (the number of collisions it must make in reaching the center each time) (Roger Blandford, private communication).

Not only is the chaotic transfer of heat long, but so is the build up of the hypernova.  It is the most energetic part of the whole explosion, reflecting better the rotational energy of the BH, the part giving rise to the X-rays having a very low efficiency because of the net ram pressure energy that must be paid to clear their way through the He star.

Thus, the $\sim6$ bethes of the hypernova energy should be doubled, assuming equipartition between kinetic and heat energies, as our estimate of rotational energy delivered to the binary.  This is perhaps $\sim1/2$ of the total rotational energy, the BH being kept rotating following the explosion.  The total energy is, therefore, $\sim40\%$ of that of LMC X$-$3 \citep{Bro08}, which was estimated to be that of a cosmological GRB.

\section{A Detailed Comparison of Core Masses}\label{Core}

It is clear from the amount of $^{56}$Ni production that the BH in SN1987A and the one in SN2008D are nearly equal in mass. Both \citet{Bet95} and \citet{Tan08} found the compact object mass to be determined by the amount of $^{56}$Ni production. Indeed, for SN1987A the production was $7.5\%$ and for SN2008D, 7\%. We shall see that, although this is a small difference, all of the small differences favor the BH in 1987A be slightly more massive than in SN2008D.

We shall use the \citet{Woo95} massive star evolution. In their Table~3 they give the Fe core masses, for their Model A for solar metallicity. From the table we see that the Fe core of SN1987A in the \citet{Bet95} upper limit of $1.56\msun$ would come from a ZAMS $18\msun$ Fe core, and the upper limit of \citet{Tan08} of $1.8\msun$ would essentially come from a ZAMS $22\msun$ Fe core. The main point we want to make here is that the mass of the BH comes from the Fe core, just as \citet{Bet95} in reconstructing the explosion SN1987A that the Fe production determined the mass of the BH.

First of all, one difference is that SN1987A had a Type II explosion. The main effect from this in the difference we discuss is that when the shock wave hits the hydrogen envelope, there is a compression wave which moves backwards and sharpens into a reverse shock. However, this takes $\sim 1$ yr to come back to the compact object \citep{Che89}, so it has no effect and the H-envelope does not have an effect on the compact object.

Secondly, the SN2008D progenitor is in strong rotation. 
This means that the neutron star is held up with a somewhat higher mass than it would have without the rotation by the centrifugal force before dropping into the BH. Additionally, the rotating compact object in SN2008D is heated by the viscosity from the interacting BH which means that it should produce more Fe than in SN1987A.

The $22\msun$ for SN2008D arrived at by \citet{Tan08} is motivated by the need of $6\msun$ of He by the $v_{\rm ph}=7,500$ km s$^{-1}$ of the photosphere, in the He layer, would only be $<3500$ km s$^{-1}$ were the He $4\msun$. However, they mentioned that {\it ``the C-abundance in the He layer is poorly known because of the uncertainties involved in the C-production by convective  3 $\alpha$-reaction in progenitor models and those in the Rayleigh-Taylor instability at the He/(C$+$O) interface during explosions, which tends to be stronger for lower mass He stars."} \citet{Tan08} find that some enhancement of C is needed in their $4\msun$ and $6\msun$ in the He layer {\it ``to reproduce the observed light curve near the peak more nicely."}

The $^{12}$C burning was carried out very carefully by Alex Heger \citep{Heg01}, and reported in \citet{Bro01}. The threshold for BH formation is at $\sim 18\msun$, just at the ZAMS mass of SN1987A, just before the central carbon abundance drops below $15\%$, the necessary amount for convective carbon burning. From $\gsim 18 \msun$ to $\sim 24\msun$ there is not enough $^{12}$C for convective burning; the $^{12}$C curves come from shell burning, which is somewhat chaotic and which takes place at  temperature much higher than the convective burning. The abundance of central $^{12}$C is $\sim 10\%$, not  lot less than necessary for convective burning.

This component of high energy $^{12}$C shell burning has not been included in \citet{Tan08}. Looking at \citet{Tan08}'s Fig.~2 light curve, it could easily convert a $4.5\msun$ He light curve for an $18\msun$ progenitor into that of a $22\msun$ progenitor.

We find, then, because of all of the reasons given above, that the progenitor of SN2008D should be less massive than that of SN1987A, although the difference in mass is not large. At the moment, a good estimate would be the \citet{Bet95} $1.5\msun$. We knew that the most massive well measured neutron star is the Hulse-Taylor $1.44\msun$ pulsar. Error bars are such that there is no good case for a neutron star more massive than this. This leaves us with the \citet{Bet95} $1.5\msun$ as maximum neutron star mass. This mass has been assaulted
several times but the relevant observations of J0751$+$1807 in our Galaxy \citep{Nic05} have been withdrawn \citep{Nic07,Bro08b}.

\section{Relation to Other GRBs}\label{GRBs}

XRF080109 is subluminous, of the type described by \citet{Bro07}.  It is quite different from the long GRBs discussed by \citet{Fru06} which were in irregular dwarf star-forming galaxies of low metallicity.  As discussed by \citet{Bro07} and \citet{Bro08}, the low metallicity is phenomenological; the real dynamics is in the donor (secondary star) which must be just right, $\sim(4-5)\msun$ in order to have a GRB of cosmological energy.  The dynamics come in in the mass dependence on this metallicity.

\citet{Woo86} suggested that stars heavier than $\sim25\msun$ goes to more massive BHs, also that such stars first explode, exhibiting light curves of Type II supernovae and returning matter to the galaxy, and then collapse into BHs.  The compact core is, for a certain range of masses, stabilized by thermal pressure during the period of Kelvin-Helmholtz contraction long enough to carry out nucleosynthesis, going into a BH after cooling and deleptonization.  This was certainly true for SN1987A, so we move Woosley and Weaver's $25\msun$ down to $18\msun$.
Presently we need BHs from ZAMS mass $\sim30\msun$ to fall directly into BHs.  Probably in between they return matter to the Galaxy and then go into BHs.

SN2008D definitely was in the class of delayed explosions.  A weak X-ray flash (XRF080109) was detected early, only $10.5$ hours after the SWIFT detection of a Type I$_{c}$ supernova explosion.  A first dim maximum in the light curve was reached less than 2 days after the XRF.  After a brief decline, the luminosity increased again, reaching principal maximum $\sim $ 19 days after the XRF \citep{Maz08}.

The long GRBs discussed by \citet{Fru06} are similar to LMC X$-$3, coming from star-forming irregular dwarf galaxies involving more massive progenitors.  They are, therefore, greatly different from the XRFs discussed in this paper.

\section{Conclusions}\label{Concl}

We conclude that SN2008D which was luminous and of which we could see the very beginning with He shell burning left a BH core which powered the XRF by its rotational energy and powered the supernova by viscous heating of the remainder of the He star following the collapse of its inner part into the BH.  We have been able to reconstruct the scenario from the detailed study of XRF060218/SN2006aj, in which the XRF progenitor had about the same mass, but the donor (secondary) star was about double as massive.

\section*{Acknowledgments}
G.E.B. was supported by the US Department of Energy under Grant No. DE-FG02-88ER40388.
C.H.L. was supported by the BAERI Nuclear R\&D program (M20808740002) of MEST/KOSEF.



\end{document}